\documentstyle[prl,aps,twocolumn,epsf]{revtex}
\begin{document}

\title{{\bf On the origin of reentrance in 2D Josephson
Junction Arrays}}
\author{F.M. Araujo-Moreira,$^{1}$ W. Maluf,$^{1}$ and S. Sergeenkov$^{1,2}$}
\address{$^{1}$Departamento de F\'{i}sica e Engenharia\\
F\'{i}sica, Grupo de Materiais e Dispositivos, Centro\\
Multidisciplinar para o Desenvolvimento de Materiais Cer\^amicos,\\
Universidade Federal de S\~ao Carlos, S\~ao Carlos, SP, 13565-905 Brazil\\
$^{2}$ Bogoliubov Laboratory of Theoretical Physics, Joint
Institute for Nuclear Research,\\ Dubna 141980, Moscow Region,
Russia }
\date{\today}
\maketitle
A comparative study of the magnetic properties of shunted and
unshunted two-dimensional Josephson junction arrays (2D-JJA) is
presented. Using a single-plaquette approximation of the 2D-JJA
model, we were able to successfully fit all our experimental data
(for the temperature, AC and DC field dependencies of
susceptibility) and demonstrate that the dynamic reentrance of AC
susceptibility is directly linked to the value of the
Stewart-McCumber parameter $\beta _{C}$. Based on extensive
numerical simulations, a phase diagram $\beta _{C}-\beta _{L}$ is
plotted which demarcates the border between the reentrant and
non-reentrant behavior in the arrays.


\narrowtext

\section{Introduction}

According to the current paradigm, paramagnetic Meissner effect
(PME)~\cite {1,2,3,4,5,6} can be related to the presence of $\pi
$-junctions~\cite{7}, either resulting from the presence of
magnetic impurities in the junction~\cite{8,9} or from
unconventional pairing symmetry~\cite{10}. Other possible
explanations of this phenomenon are based on flux
trapping~\cite{11} and flux compression effects~\cite{12}
including also an important role of the surface of the
sample~\cite{3}. Besides, in the experiments with unshunted
2D-JJA, we have previously reported~\cite{13} that PME manifests
itself through a dynamic reentrance (DR) of the AC magnetic
susceptibility as a function of temperature. These results have
been further corroborated by Nielsen et al.~\cite{14} and De Leo
et al.~ \cite{15} who argued that PME can be simply related to
magnetic screening in multiply connected superconductors. So, the
main question is: which parameters are directly responsible for
the presence (or absence) of DR in artificially prepared arrays?

Previously (also within the single plaquette approximation),
Barbara et al.~\cite{13} have briefly discussed the effects of
varying $\beta _{L}$ on the observed dynamic reentrance with the
main emphasis on the behavior of 2D-JJA samples with high (and
fixed) values of $\beta _{C}$. However, to our knowledge, up to
date no systematic study (either experimental or theoretical) has
been done on how the $\beta _{C}$ value itself affects the
reentrance behavior. In the present work, by a comparative study
of the magnetic properties of shunted and unshunted 2D-JJA, we
propose an answer to this open question. Namely, by using
experimental and theoretical results, we will demonstrate that
only arrays with sufficiently large value of the Stewart-McCumber
parameter $\beta _{C}$ will exhibit the dynamic reentrance
behavior (and hence PME).

\section{Experimental Results }

To measure the complex AC susceptibility in our arrays we used a
high-sensitive home-made susceptometer based on the so-called
screening method in the reflection configuration~\cite{16,17,18}.
The experimental system was calibrated by using a high-quality
niobium thin film. Previously~\cite{18}, we have shown that the
calibrated output complex voltage of the experimental setup
corresponds to the complex AC susceptibility.

To experimentally investigate the origin of the reentrance, we
have measured $\chi ^{\prime }(T)$ for three sets of shunted and
unshunted samples obtained from different makers (Westinghouse and
Hypress) under the same conditions of the amplitude of the
excitation field $h_{ac}$ ($1mOe<h_{ac}<10Oe$), external magnetic
field $H_{dc}$ ($0<H_{dc}<500Oe$) parallel to the plane of the
sample, and frequency of AC field $\omega =2\pi f$ (fixed at
$f=20kHz$). Unshunted 2D-JJAs are formed by loops of niobium
islands linked through $Nb-AlO_{x}-Nb$ Josephson junctions while
shunted 2D-JJAs have a molybdenum shunt resistor (with
$R_{sh}\simeq 2.2\Omega $) short-circuiting each junction (see
Fig.1). Both shunted and unshunted samples have rectangular
geometry and consist of $100\times 150$ tunnel junctions. The unit
cell for both types of arrays has square geometry with lattice
spacing $a\simeq 46\mu m$ and a single junction area of $5\times
5\mu m^{2}$. The
critical current density for the junctions forming the arrays is about $%
600A/cm^{2}$ at $4.2K$. Besides, for the unshunted samples $\beta
_{C}(4.2K)\simeq 30$ and $\beta _{L}(4.2K)\simeq 30$, while for
shunted samples $\beta _{C}(4.2K)\simeq 1$ and $\beta _{L}(4.2K)\simeq 30$
where~\cite{19} $\beta _{L}(T)=\frac{%
2\pi LI_{C}(T)}{\Phi _{0}}$ and $\beta _{C}(T)=\frac{2\pi
C_{J}R_{J}^{2}I_{C}(T)}{\Phi _{0}}$. Here, $C_{J}\simeq 0.58pF$ is
the capacitance, $R_{J}\simeq 10.4\Omega $ the quasi-particle
resistance (of unshunted array), and $I_{C}(4.2K)\simeq 150\mu A$
critical current of the Josephson junction. $\Phi _{0}$ is the
quantum of magnetic flux. The parameter $\beta _{L}$ is
proportional to the number of flux quanta that can be screened by
the maximum critical current in the junctions, while the
Stewart-McCumber parameter $\beta _{C}$ basically reflects the
quality of the junctions in arrays.

It is well established that both magnetic and transport properties
of any superconducting material can be described via a
two-component response~\cite{20}, the {\it intragranular}
(associated with the grains exhibiting bulk superconducting
properties) and {\it intergranular} (associated with weak-link
structure) contributions~\cite{21,22}. Likewise, artificially
prepared JJAs (consisting of superconducting islands, arranged in
a symmetrical periodic lattice and coupled by Josephson junctions)
will produce a similar response~\cite{23}.

Since our shunted and unshunted samples have the same value of
$\beta _{L}$ and different values of $\beta _{C}$, it is possible
to verify the dependence of the reentrance effect on the value of
the Stewart-McCumber parameter. For the unshunted 2D-JJA (Fig. 2a)
we have found that for an AC field lower than $50mOe$ (when the
array is in the Meissner-like state) the behavior of $\chi
^{\prime }(T)$ is quite similar to homogeneous superconducting
samples, while for $h_{ac}>50mOe$ (when the array is in the
mixed-like state with practically homogeneous flux distribution)
these samples exhibit a clear reentrant behavior of
susceptibility~\cite{13}. At the same time, the identical
experiments performed on the shunted samples produced no evidence
of any reentrance for all values of $h_{ac}$ (see Fig. 2b). It is
important to point out that the analysis of the experimentally
obtained imaginary component of susceptibility $\chi ^{\prime
\prime }(T)$ shows that for the highest AC magnetic field
amplitudes (of about $200mOe$) dissipation remains small. Namely,
for typical values of the AC amplitude, $h_{ac}=100mOe$ (which
corresponds to about $10$ vortices per unit cell) the imaginary
component is about $15$ times smaller than its real counterpart.
Hence contribution from the dissipation of vortices to the
observed phenomena can be safely neglected.

To further study this unexpected behavior we have also performed
experiments where we measure $\chi ^{\prime }(T)$ for different
values of $H_{dc}$ keeping the value of $h_{ac}$ constant. The
influence of DC fields on reentrance in unshunted samples is shown
in Fig. 3. On the other hand, the shunted samples still show no
signs of reentrance, following a familiar pattern of field-induced
gradual diminishing of superconducting phase (very similar to a
zero DC field flat-like behavior seen in Fig.2b).

To understand the influence of DC field on reentrance observed in
unshunted arrays, it is important to emphasize that for our sample
geometry this parallel field suppresses the critical current
$I_{C}$ of each junction without introducing any detectable flux
into the plaquettes of the array. Thus, a parallel DC magnetic
field allows us to vary $I_{C}$ independently from temperature
and/or applied perpendicular AC field. The measurements show (see
Fig.3) that the position of the reentrance is tuned by $H_{dc}$.
We also observe that the value of temperature $T_{min}$ (at which
$\chi ^{\prime }(T)$ has a minimum) first shifts towards lower
temperatures as we raise $H_{dc}$ (for small DC fields) and then
bounces back (for higher values of $H_{dc}$). This non-monotonic
behavior is consistent with the weakening of $I_C(T)$ and
corresponds to Fraunhofer-like dependence of the Josephson
junction critical current on DC magnetic field applied in the
plane of the junction. We measured $I_{C}$ from transport
current-voltage characteristics, at different values of $H_{dc}$
at $T=4.2K$  and found that $\chi ^{\prime }(T=4.2K)$, obtained
from the isotherm $T=4.2K$ (similar to that given in Fig.3), shows
the same Fraunhofer-like dependence on $H_{dc}$ as the critical
current $I_{C}(H_{dc})$ of the junctions forming the array (see
Fig.4). This gives further proof that only the junction critical
current is varied in this experiment. This also indicates that the
screening currents at low temperature (i.e., in the reentrant
region) are proportional to the critical currents of the
junctions. In addition, this shows an alternative way to obtain
$I_{C}(H_{dc})$ dependence in big arrays. And finally, a sharp
Fraunhofer-like pattern observed in both arrays clearly reflects a
rather strong coherence (with negligible distribution of critical
currents and sizes of the individual junctions) which is based on
highly correlated response of {\it all} single junctions forming
the arrays, thus proving their high quality. Such a unique
behavior of Josephson junctions in our samples provides a
necessary justification for suggested theoretical interpretation
of the obtained experimental results. Namely, based on the
above-mentioned properties of our arrays, we have found that
practically all the experimental results can be explained by
analyzing the dynamics of just a single unit cell in the array.

\section{Theoretical Interpretation and Numerical Simulations}

To understand the different behavior of the AC susceptibility
observed in shunted and unshunted 2D-JJAs, in principle one would
need to analyze in detail the flux dynamics in these arrays.
However, as we have previously reported~\cite{13}, because of the
well-defined periodic structure of our arrays (with no visible
distribution of junction sizes and critical currents), it is
reasonable to expect that the experimental results obtained from
the magnetic properties of our 2D-JJAs can be quite satisfactory
explained by analyzing the dynamics of a single unit cell
(plaquette) of the array. An excellent agreement between a
single-loop approximation and the observed behavior (seen through
the data fits) justifies {\it a posteriori} our assumption. It is
important to mention that the idea to use a single unit cell to
qualitatively understand PME was first suggested by Auletta et
al.~\cite{24}. They simulated the field-cooled DC magnetic
susceptibility of a single-junction loop and found a paramagnetic
signal at low values of external magnetic field.

In our calculations and numerical simulations, the unit cell is a
loop containing four identical Josephson junctions and the
measurements correspond to the zero-field cooling (ZFC) AC
magnetic susceptibility. We consider the junctions of the single
unit cell as having capacitance $C_{J}$, quasi-particle resistance
$R_{J}$ and critical current $I_{C}$. We have used this simple
four-junctions model to study the magnetic behavior of our 2D-JJA
by calculating the AC complex magnetic susceptibility $\chi =\chi
^{\prime }+i\chi ^{\prime \prime }$ as a function of $T$, $\beta
_{C}$ and $\beta _{L}$. Specifically, shunted samples are
identified through low values of the McCumber parameter ($\beta
_{C}\approx 1$) while high values ($\beta _{C}\gg 1$) indicate an
unshunted 2D-JJA.

If we apply an AC external field $B_{ac}(t)=\mu _0h_{ac}\cos
\omega t$ normally to the 2D-JJA and a DC field $B_{dc}=\mu
_0H_{dc}$ parallel to the array, then the total magnetic flux
$\Phi (t)$ threading the four-junction superconducting loop is
given by $\Phi (t)=\Phi _{ext}(t)+LI(t)$ where $L$ is the loop
inductance, $\Phi _{ext}(t)=SB_{ac}(t)+ldB_{dc}$ is the flux
related to the applied magnetic field (with $l\times d$ being the
size of the single junction area, and $S\simeq a^2$ being the
projected area of the loop), and the circulating current in the
loop reads
\begin{equation}
I(t)=I_C(T)\sin \phi _i(t)+\frac{\Phi _0}{2\pi R_J}\frac{d\phi
_i}{dt}+ \frac{C_J\Phi _0}{2\pi}\frac{d^2\phi _i}{dt^2}
\end{equation}
Here $\phi _i(t)$ is the gauge-invariant superconducting phase difference
across the $i$th junction, and $\Phi _0$ is the magnetic flux quantum.

Since the inductance of each loop is $L=\mu _{0}a \simeq 64pH$ and
the critical current of each junction is $I_{C}\simeq 150\mu A$,
for the mixed-state region (above $50mOe$) we can safely neglect
the self-field effects because in this region $LI(t)$ is always
smaller than $\Phi _{ext}(t)$. Besides, since the length $l$ and
the width $w$ of each junction in our array is smaller than the
Josephson penetration depth $\lambda _{J}=\sqrt{\Phi _{0}/2\pi \mu
_{0}dj_{c0}}$ (where $j_{c0}$ is the critical current density of
the junction, and $d=2\lambda _{L}+\xi $ is the size of the
contact area with $\lambda _{L}(T)$ being the London penetration
depth of the junction
and $\xi $ an insulator thickness), namely $l\simeq w\simeq 5\mu m$ and $%
\lambda _{J}\simeq 20\mu m$ (using $j_{c0}\simeq 600A/cm^{2}$ and
$\lambda _{L}\simeq 39nm$ for $Nb$ at $T=4.2K$), we can adopt the
small-junction approximation~\cite{19} for the gauge-invariant
superconducting phase difference across the $i$th junction (for
simplicity we assume as usual~\cite{13} that $\phi _{1}=\phi
_{2}=\phi _{3}=\phi _{4}\equiv \phi _{i}$)
\begin{equation}
\phi _{i}(t)=\phi _{0}(H_{dc})+\frac{2\pi B_{ac}(t)S}{\Phi _{0}}
\end{equation}
where $\phi _{0}(H_{dc})=\phi _{0}(0)+2\pi \mu _{0}H_{dc}dl/\Phi _{0}$ with $%
\phi _{0}(0)$ being the initial phase difference.

To properly treat the magnetic properties of the system, let us introduce
the following Hamiltonian
\begin{equation}
{\cal H}(t)=J\sum _{i=1}^4[1-\cos \phi _i(t)]+\frac{1}{2}LI^2(t)
\end{equation}
which describes the tunneling (first term) and inductive (second term)
contributions to the total energy of a single plaquette. Here, $J(T)=(\Phi
_0/2\pi )I_C(T)$ is the Josephson coupling energy.

The real part of the complex AC susceptibility is defined as
\begin{equation}
\chi ^{\prime}(T,h_{ac},H_{dc})=\frac{\partial M}{\partial h_{ac}}
\end{equation}
where
\begin{equation}
M(T,h_{ac},H_{dc})=-\frac{1}{V}\left <\frac{\partial {\cal H}}{\partial
h_{ac}}\right >
\end{equation}
is the net magnetization of the plaquette. Here $V$ is the
sample's volume, and $<...>$ denotes the time averaging over the
period $2\pi /\omega $, namely
\begin{equation}
<A>=\frac{1}{2\pi}\int_0^{2\pi}d(\omega t)A(t)
\end{equation}
Taking into account the well-known~\cite{25} analytical
approximation of the BCS gap parameter (valid for all
temperatures), $\Delta (T)=\Delta (0)\tanh \left(
2.2\sqrt{\frac{T_{c}-T}{T}}\right)$ for the explicit temperature
dependence of the Josephson critical current
\begin{equation}
I_{C}(T)=I_{C}(0)\left[ \frac{\Delta (T)}{\Delta (0)}\right] \tanh
\left[ \frac{\Delta (T)}{2k_{B}T}\right]
\end{equation}
we successfully fitted all our data using the following set of
parameters: $\phi _{0}(0)=\frac{\pi}{2}$ (which corresponds to the
maximum Josephson current within a plaquette), $\beta _{L}(0)=32$,
$\beta _{C}(0)=32$ (for unshunted array) and $\beta _{C}(0)=1.2$
(for shunted array). The corresponding fits are shown by solid
lines in Figs.2 and 3 for the experimental values of AC and DC
field amplitudes.

In the mixed-state region and for low enough frequencies (this
assumption is well-satisfied because in our case $\omega \ll
\omega _{LR}$ and $\omega \ll \omega _{LC}$ where $\omega
_{LR}=R/L$ and $\omega _{LC}=1/\sqrt{LC}$ are the two
characteristic frequencies of the problem) from Eqs.(3)-(6) we
obtain the following approximate analytical expression for the
susceptibility of the plaquette
\begin{eqnarray}
\chi ^{\prime}(T,h_{ac},H_{dc})& \simeq &-\chi _0(T) [\beta
_L(T)f_1(b) \cos \left (\frac{2H_{dc}}{H_0}\right ) \\ \nonumber
&+&f_2(b)\sin \left (\frac{H_{dc}}{H_0}\right )-\beta _C^{-1}(T)]
\end{eqnarray}
where $\chi _0(T)=\pi S^2I_C(T)/V\Phi_0$, $H_0=\Phi
_{0}/(2\pi \mu _{0}dl) \simeq 10 Oe$, $f_1(b)=J_0(2b)-J_2(2b)$, and $%
f_2(b)=J_0(b)-bJ_1(b)-3J_2(b)+bJ_3(b)$ with $b=2\pi S\mu
_0h_{ac}/\Phi _0$ and $J_n(x)$ being the Bessel function of the
$n$th order.

Notice also that the analysis of Eq.(8) reproduces the observed
Fraunhofer-like behavior of the susceptibility in applied DC field
(see Fig.4) and the above-mentioned fine tuning of the reentrance
effect (see also Ref.13). Indeed, according to Eq.(8) (and in
agreement with the observations), for small DC fields the minimum
temperature $T_{min}$ (indicating the beginning of the reentrant
transition) varies with $H_{dc}$ as follows, $1-T_{min}/T_C \simeq
H_{dc}/H_0$.

To further test our interpretation and verify the influence of the
parameter $\beta _{C}$ on the reentrance, we have also performed
extensive numerical simulations of the four-junction model
previously described but without a simplifying assumption about
the explicit form of the phase difference based on Eq.(2). More
precisely, we obtained the temperature behavior of the
susceptibility by solving the set of equations responsible for the
flux dynamics within a single plaquette and based on Eq.(1) for
the total current $I(t)$, the equation for the total flux $\Phi
(t)=\Phi _{ext}(t)+LI(t)$ and the flux quantization condition for
four junctions, namely $\phi _i(t)=\frac{\pi
}{2}\left(n+\frac{\Phi }{\Phi _{0}}\right )$ where $n$ is an
integer. Both Euler and fourth-order Runge-Kutta integration
methods provided the same numerical results. In Fig.5 we show the
real component of the simulated susceptibility $\chi (T)$
corresponding to the fixed value of $\beta _{C}(T=4.2K)=1$
(shunted samples) and different values of $\beta _{L}(T=4.2K)=1$,
$10$, $15$, $20$, $30$, $40$, $50$, $60$, $90$, $150$ and $200$.
As expected, for this low value of $\beta _{C}$ reentrance is not
observed for any values of $\beta _{L}$. On the other hand, Fig.6
shows the real component of the simulated $\chi (T)$ but now using
fixed value of $\beta _{L}(T=4.2K)=30$ and different values of
$\beta _{C}(T=4.2K)=1$, $2$, $5$, $10$, $20$, $30$ and $100$. This
figure clearly shows that reentrance appears for values of $\beta
_{C}>20$. In both cases we used $h_{ac}=70mOe$. We have also
simulated the curve for shunted ($\beta _{L}=30$, $\beta _{C}=1$)
and unshunted ($\beta _{L}=30$, $\beta _{C}=30$) samples for
different values of $h_{ac}$ (see Fig.7). In this case the values
of the parameters $\beta _{L}$ and $\beta _{C}$ were chosen from
our real 2D-JJA samples. Again, our simulations confirm that
dynamic reentrance does not occur for low values of $\beta _{C}$,
independently of the values of $\beta _{L}$ and $h_{ac}$.

The following comment is in order regarding some irregularities
visibly seen in Figs.(5)-(7) around the transition regions from
non-reentrant to reentrant behavior. It is important to emphasize
that the above irregularities are just artifacts of the numerical
simulations due to the conventional slow-converging real-time
reiteration procedures~\cite{13}. They neither correspond to any
experimentally observed behavior (within the accuracy of the
measurements technique and data acquisition), nor they reflect any
irregular features of the considered here theoretical model (which
predicts a smooth temperature dependence seen through the data
fits). As usual, to avoid this kind of artificial (non-physical)
discontinuity, more powerful computers are needed.

Based on the above extensive numerical simulations, a resulting
{\it phase diagram} $\beta _{C}-\beta _{L}$ (taken for $T=1K$,
$h_{ac}=70mOe$, and $H_{dc}=0$) is depicted in Fig.8 which clearly
demarcates the border between the reentrant (white area) and
non-reentrant (shaded area) behavior in the arrays for different
values of $\beta _{L}(T)$ and $\beta _{C}(T)$ parameters at given
temperature. In other words, if $\beta _{L}$ and $\beta _{C}$
parameters of any realistic array have the values inside the white
area, this array will exhibit a reentrant behavior. In addition,
this diagram shows that one can prepare a reentrance exhibiting
array by changing one of the parameters (usually, it is much
easier to change $\beta _{C}$ by tuning the shunt resistance
rather than the geometry related inductance parameter $\beta
_{L}$).

It is instructive to mention that a hyperbolic-like character of
$\beta _{L}$ vs $\beta _{C}$ law (seen in Fig.8) is virtually
present in the approximate analytical expression for the
susceptibility of the plaquette given by Eq.(8) (notice however
that this expression can not be used to produce any quantitative
prediction because the neglected in Eq.(8) frequency-related terms
depend on $\beta _{L}$ and $\beta _{C}$ parameters as well). A
qualitative behavior of the envelope of the {\it phase diagram}
(depicted in Fig.8) with DC magnetic field $H_{dc}$ (for $T=1K$
and $h_{ac}=70mOe$), obtained using Eq.(8), is shown in Fig.9.

And finally, to understand how small values of $\beta _{C}$
parameter affect the flux dynamics in shunted arrays, we have
analyzed the $\Phi _{tot}$($\Phi _{ext}$) diagram. Similarly to
those results previously obtained from unshunted
samples~\cite{13}, for a shunted sample at fixed temperature this
curve is also very hysteretic (see Fig.10). In both cases, $\Phi
_{tot}$ vs. $\Phi _{ext}$ shows multiple branches intersecting the
line $\Phi _{tot}=0$ which corresponds to diamagnetic states. For
all the other branches, the intersection with the line $\Phi
_{tot}=\Phi _{ext}$ corresponds to the boundary between
diamagnetic states (negative values of $ \chi ^{\prime }$) and
paramagnetic states (positive values of $\chi ^{\prime }$). As we
have reported before~\cite{13}, for unshunted 2D-JJA at
temperatures below $7.6K$ the appearance of the first and third
branches adds a paramagnetic contribution to the average value of
$\chi ^{\prime }$. When $\beta _{C}$ is small (shunted arrays),
the analysis of these curves shows that there is no reentrance at
low temperatures because in this case the second branch appears to
be energetically stable, giving an extra diamagnetic contribution
which overwhelms the paramagnetic contribution from subsequent
branches. In other words, for low enough values of $\beta _{C}$
(when the samples are ZFC and then measured at small values of the
magnetic field), most of the loops will be in the diamagnetic
states, and no paramagnetic response is registered. As a result,
the flux quanta cannot get trapped into the loops even by the
following field-cooling process in small values of the magnetic
field. In this case the superconducting phases and the junctions
will have the same diamagnetic response and the resulting measured
value of the magnetic susceptibility will be negative (i.e.,
diamagnetic) as well. On the other hand, when $\beta _{C}$ is
large enough (unshunted arrays), the second branch becomes
energetically unstable, and the average response of the sample at
low temperatures is paramagnetic (Cf. Fig.7 from Ref.~\cite{13}).

In conclusion, our experimental and theoretical results have
demonstrated that the reentrance phenomenon (and concomitant PME)
in artificially prepared Josephson Junction Arrays is related to
the damping effects associated with the Stewart-McCumber parameter
$\beta _C$. Namely, reentrant behavior of AC susceptibility takes
place in the underdamped (unshunted) array (with large enough
value of $\beta _C$) and totally disappears in overdamped
(shunted) arrays.

\acknowledgments

We thank P. Barbara, C.J. Lobb, R.S. Newrock, and A. Sanchez for
useful discussions. S.S. and F.M.A.M. gratefully acknowledge
financial support from Brazilian Agency FAPESP under grant
03/00296-5.

\begin{figure}[tbh]
\epsfxsize=8.5cm \centerline{\hspace{0mm} \epsffile{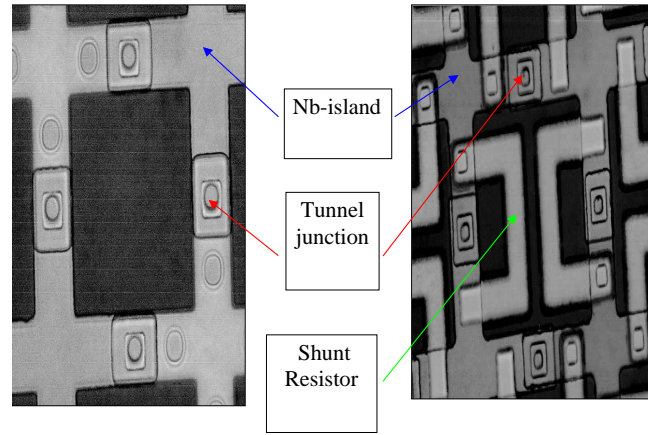}}
\vspace{0.5cm} \caption{Left: photograph of the unshunted array;
right: photograph of the shunted array. }
\end{figure}
\begin{figure}[htb]
\epsfxsize=6.5cm \centerline{\hspace{0mm} \epsffile{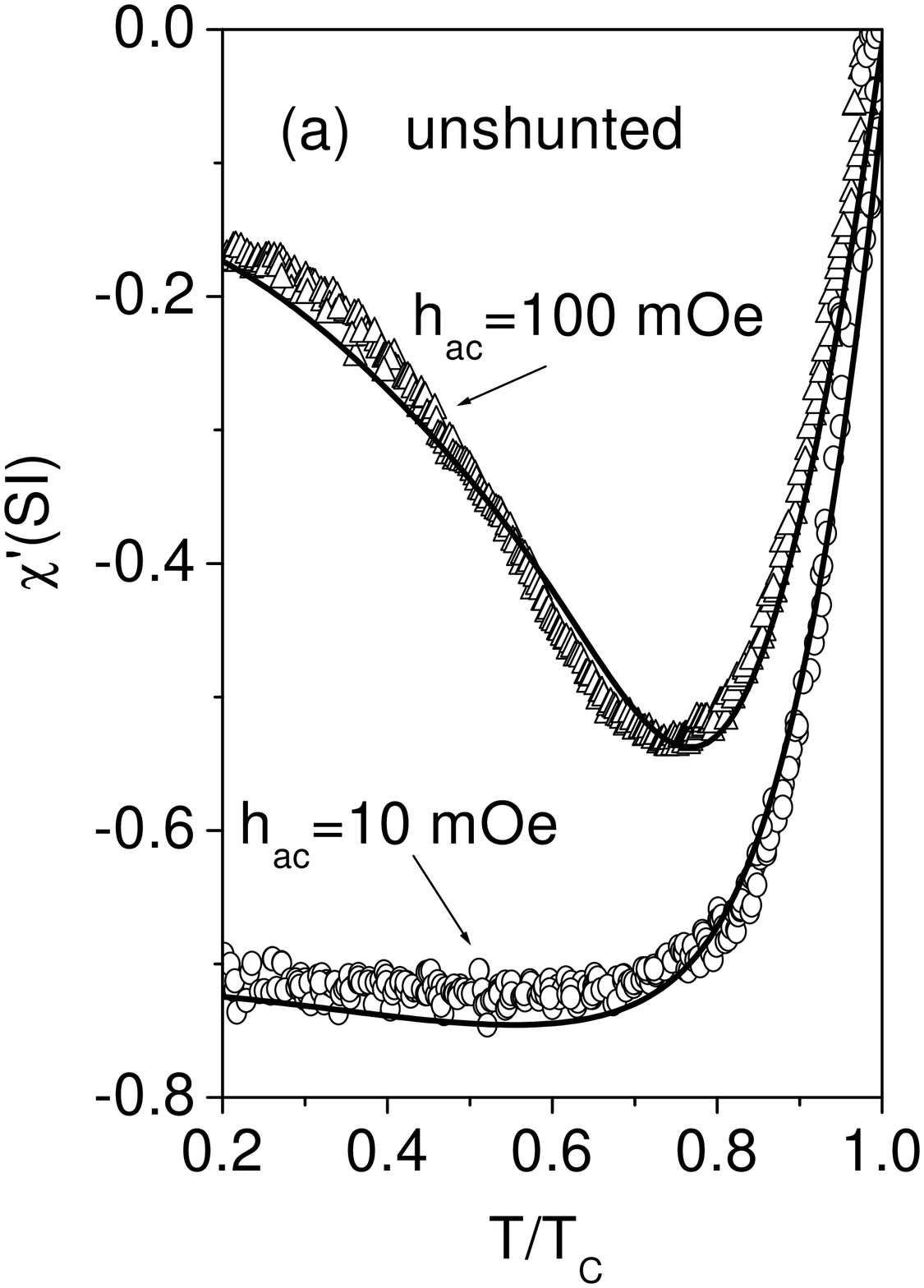}}
\vspace{0.5cm} \epsfxsize=6.5cm \centerline{\hspace{0mm}
\epsffile{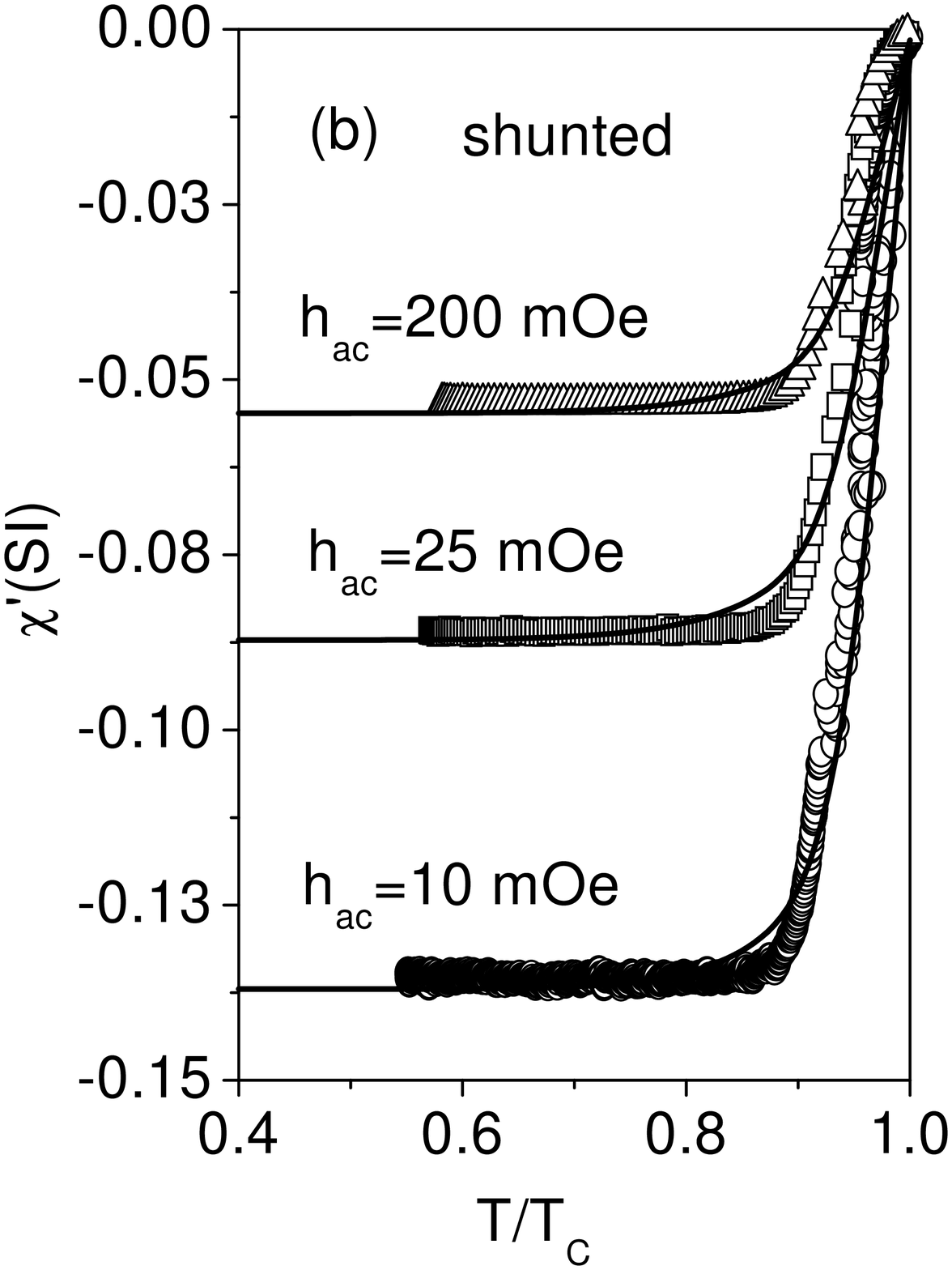}} \vspace{0.5cm} \caption{Experimental results
for $\chi '(T,h_{ac},H_{dc})$: (a) unshunted 2D-JJA for $h_{ac} =
10$ and $100 mOe$; (b) shunted 2D-JJA for  $h_{ac}=10$, $25$, and
$200 mOe$. In all these experiments $H_{dc}=0$. Solid lines are
the best fits (see text).}
\end{figure}
\begin{figure}[htb]
\epsfxsize=6.5cm \centerline{\hspace{0mm} \epsffile{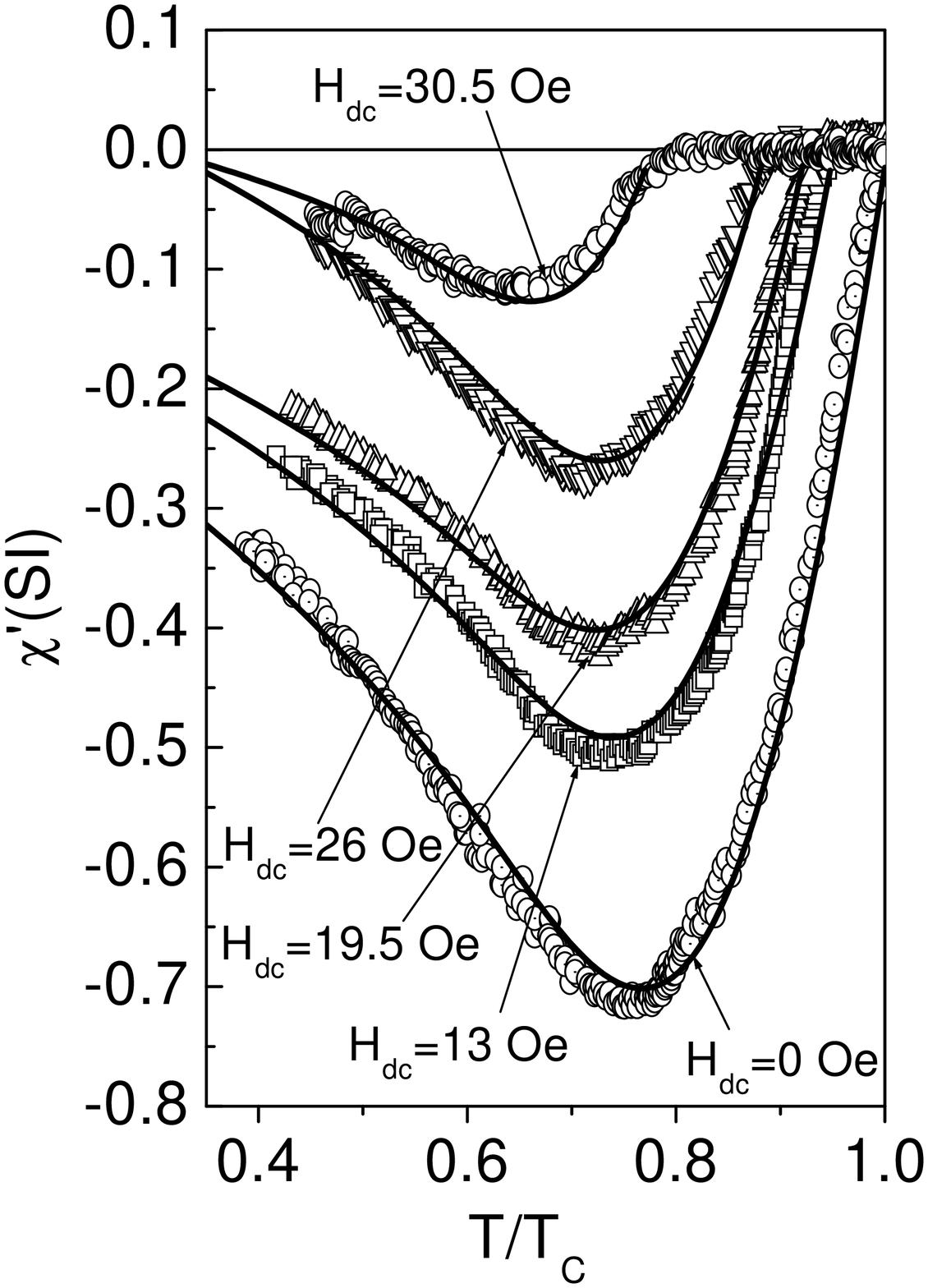}}
\vspace{0.5cm} \caption{Experimental results for $\chi
'(T,h_{ac},H_{dc})$ for unshunted 2D-JJA for $H_{dc} = 0$, $13$,
$19.5$, $26$, and $30.5 Oe$. In all these experiments  $h_{ac}=
100 mOe$. Solid lines are the best fits (see text).}
\end{figure}
\begin{figure}[htb]
\epsfxsize=8.5cm \centerline{\hspace{0mm} \epsffile{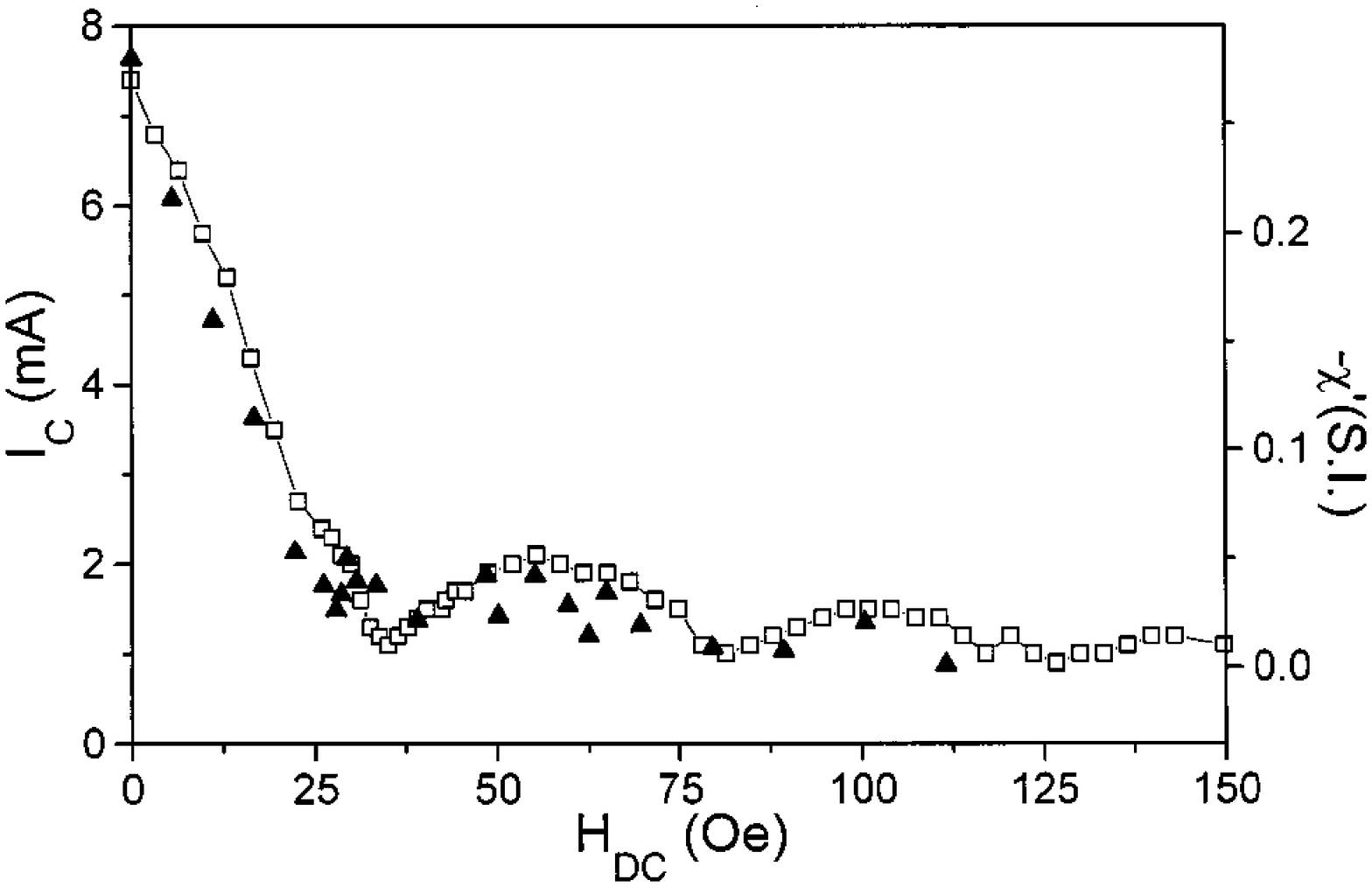}}
\vspace{0.5cm} \caption{The critical current $I_C$ (open squares)
and the real part of AC susceptibility $\chi '$ (solid triangles)
as a function of DC field $H_{dc}$ for $T=4.2K$ (from Ref.13).}
\end{figure}
\begin{figure}[htb]
\epsfxsize=6.5cm \centerline{\hspace{0mm} \epsffile{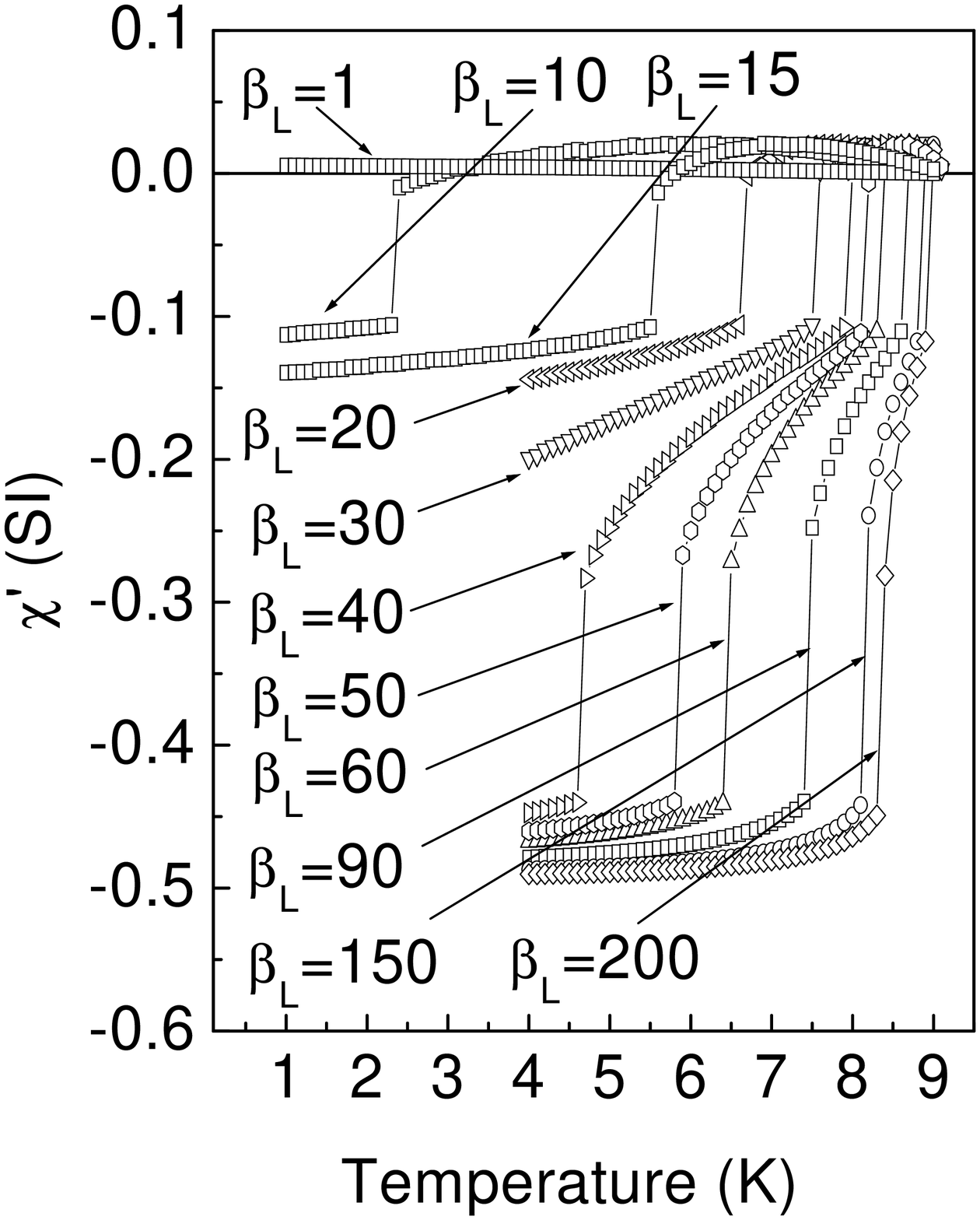}}
\vspace{0.5cm} \caption{Numerical simulation results for
$h_{ac}=70 mOe$, $H_{dc}=0$, $\beta _{C}(T=4.2K)=1$ and for
different values of $\beta _{L}(T=4.2K)$ based on Eqs.(4)-(7).}
\end{figure}
\begin{figure}[htb]
\epsfxsize=6.5cm \centerline{\hspace{0mm} \epsffile{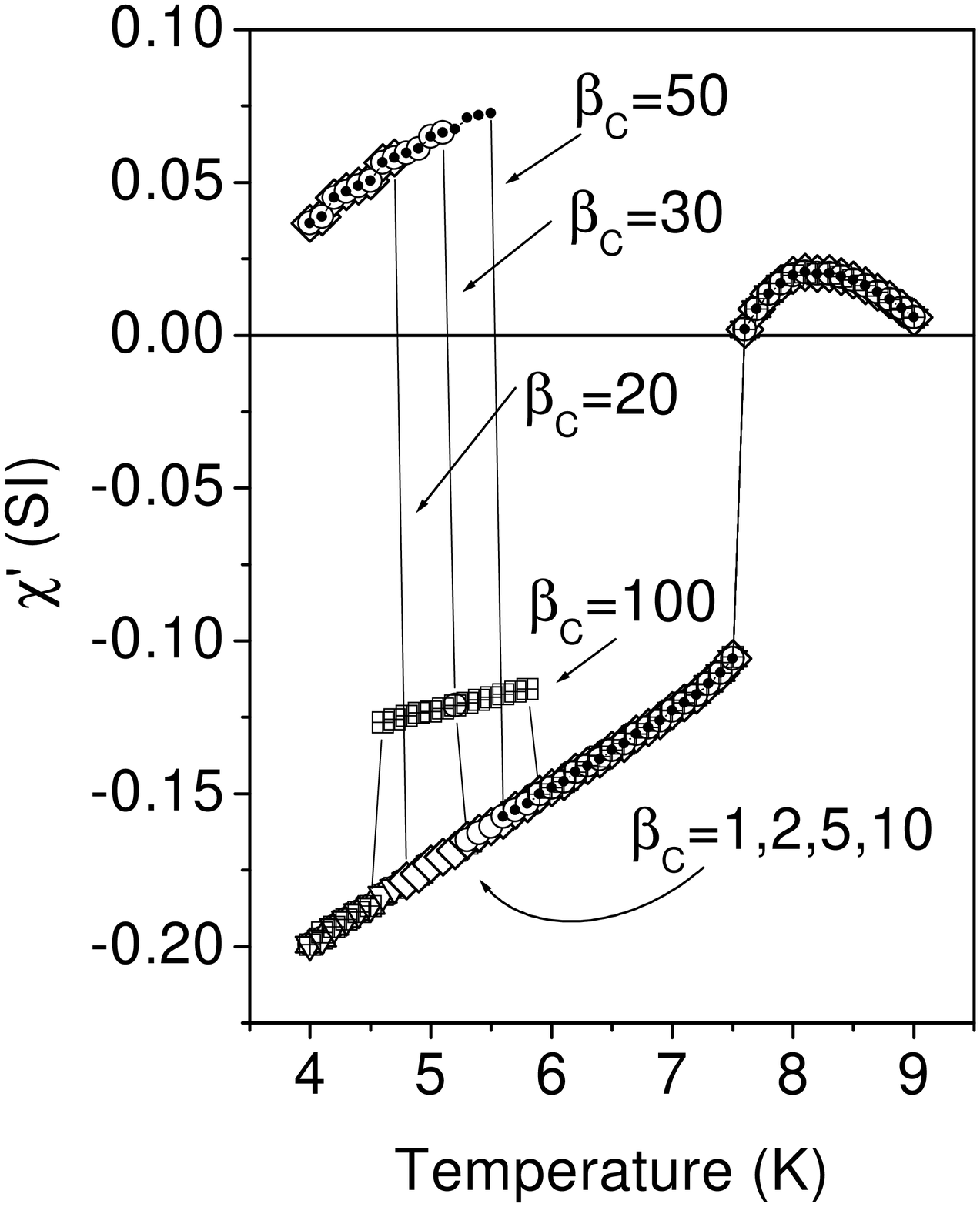}}
\vspace{0.5cm} \caption{Numerical simulation results for $h_{ac}=
70 mOe$, $H_{dc}=0$, $\beta _{L}(T=4.2K)=30$ and for different
values of $\beta _{C}(T=4.2K)$ based on Eqs.(4)-(7).}
\end{figure}
\begin{figure}[htb]
\epsfxsize=6.5cm \centerline{\hspace{0mm} \epsffile{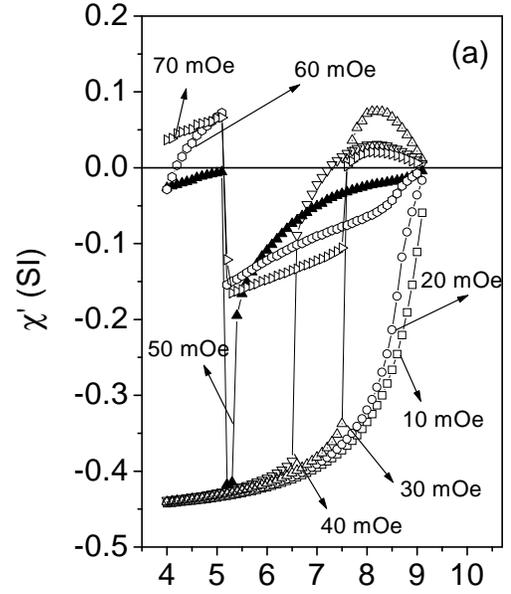}}
\vspace{0.5cm} \epsfxsize=6.5cm \centerline{\hspace{0mm}
\epsffile{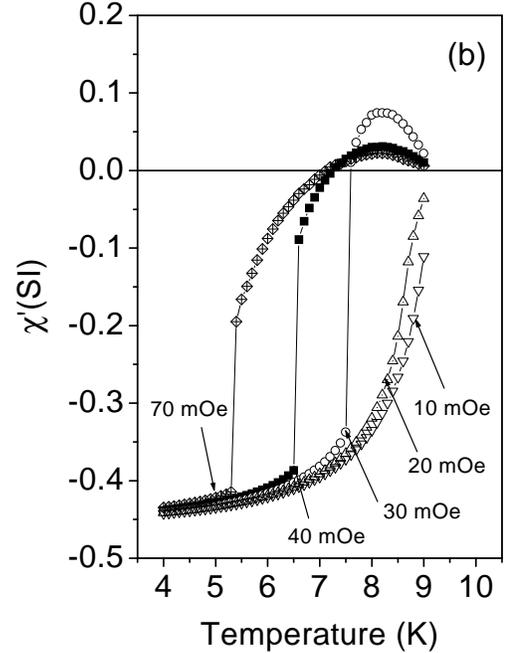}} \vspace{0.5cm} \caption{Curves of the
simulated susceptibility (for $H_{dc}=0$ and for different values
of $h_{ac}$) corresponding to (a) unshunted 2D-JJA with $\beta
_{L}(T=4.2K)=30$ and  $\beta _{C}(T=4.2K)=30$; (b) shunted 2D-JJA
with $\beta _{L}(T=4.2K)=30$ and $\beta _{C}(T=4.2K)=1$.}
\end{figure}
\begin{figure}[htb]
\epsfxsize=6.5cm \centerline{\hspace{0mm} \epsffile{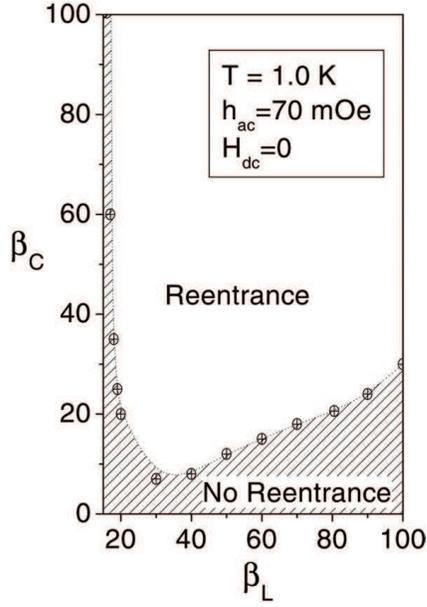}}
\vspace{0.5cm} \caption{Numerically obtained {\it phase diagram}
(taken for $T=1K$, $h_{ac}=70mOe$, and $H_{dc}=0$) which shows the
border between the reentrant (white area) and non-reentrant
(shaded area) behavior in the arrays for different values of
$\beta _{L}$ and $\beta _{C}$ parameters.}
\end{figure}
\begin{figure}[htb]
\epsfxsize=6.5cm \centerline{\hspace{0mm} \epsffile{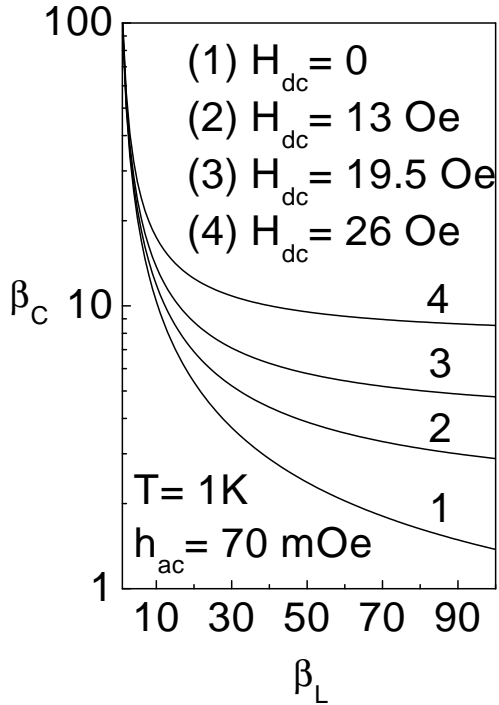}}
\vspace{0.5cm} \caption{A qualitative behavior of the envelope of
the {\it phase diagram} (shown in previous figure) with DC
magnetic field $H_{dc}$ (for $T=1K$ and $h_{ac}=70mOe$) obtained
from Eq.(8).}
\end{figure}
\begin{figure}[htb]
\epsfxsize=6.5cm \centerline{\hspace{0mm} \epsffile{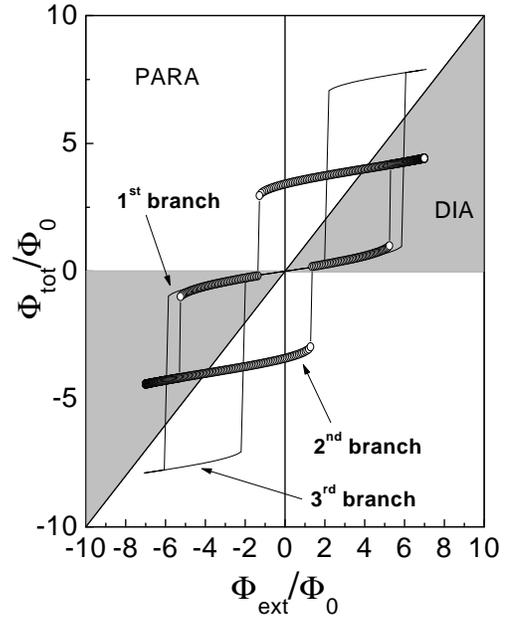}}
\vspace{0.5cm} \caption{Numerical simulation results showing $\Phi
_{tot}$ vs. $\Phi _{ext}$ for shunted 2D-JJA with $\beta
_{L}(T=4.2K)=30$ and $\beta _{C}(T=4.2K)=1$.}
\end{figure}

\end{document}